\documentclass[]{aipproc}

\usepackage{color}
\layoutstyle{6s}

\begin{document}

\title{$\rho^0$-Meson Helicity Amplitude Ratios at HERMES}

\author{Morgan J. Murray for the H{\sc ermes} Collaboration}{
	address={SUPA, School of Physics \& Astronomy, University of Glasgow, Glasgow, G12 8QQ, Scotland}
}
\keywords{lepton-nucleon scattering}
\classification{13.50.Fz,14.20,Dh}

\copyrightyear {2011}

\begin{abstract}
The study of $\rho^0$ meson helicity amplitude ratios at H{\sc ermes} shows that the amplitude hierarchy expected from pQCD is confirmed. The contribution of Unnatural Parity Exchange in the production of $\rho^0$ mesons is significant at H{\sc ermes} kinematics and there is a large phase-difference in the leading $F_{11}$ and $F_{01}$ amplitudes. The kinematic dependences of the amplitude ratios only sometimes follow theory-based expectations.
\end{abstract}

\date{\today}
\maketitle
\section{Introduction}

These proceedings report briefly on selected H{\sc ermes} results on the exclusive electroproduction of $\rho^0$ mesons on a gaseous $^1H$ or $^2H$ target~\cite{paper}.

The study of helicity amplitude ratios at H{\sc ermes} complements a previous publication on the extraction of the Spin Density Matrix Elements (SDMEs) of the same process~\cite{rhosdme}. The helicity amplitudes are extracted as ratios to the leading amplitude $F_{00}$, which describes the transition from a longitudinally polarised photon to a longitudinally polarised meson. If the ratios of the nucleon spin-flip amplitudes to $F_{00}$ were known, in addition to the ratios obtained in the present paper it would be possible to determine the amplitudes individually from the differential cross section $\textrm{d}\sigma/\textrm{d}t'$, but this knowledge is not currently available.

The previous H{\sc ermes} publication confirmed the existence of a hierarchy amongst the amplitude ratios motivated from pQCD~\cite{ivanov,kuraev} of the form $| T_{00} |^2 \approx |T_{11}|^2>| U_{11} |^2 > | T_{01} |^2 >> | T_{10} |^2...$ where any amplitude $F_{\lambda_V\lambda_\gamma}$ is the sum of an natural parity exchange amplitude ($T$) and an unnatural parity exchange amplitude ($U$). The first index $\lambda_v$ is the helicity of the produced meson and the second index $\lambda_\gamma$ is the helicity of the impinging photon. In this document we mainly consider amplitudes relating to Nucleon Helicity Conservation (NHC); natural parity exchange contributions are assumed to be dominated by amplitudes involving NHC whereas unnatural parity exchange amplitudes receive contributions from all nucleon helicity states.

\begin{center}
\begin{figure}
\includegraphics[width=0.35\textwidth]{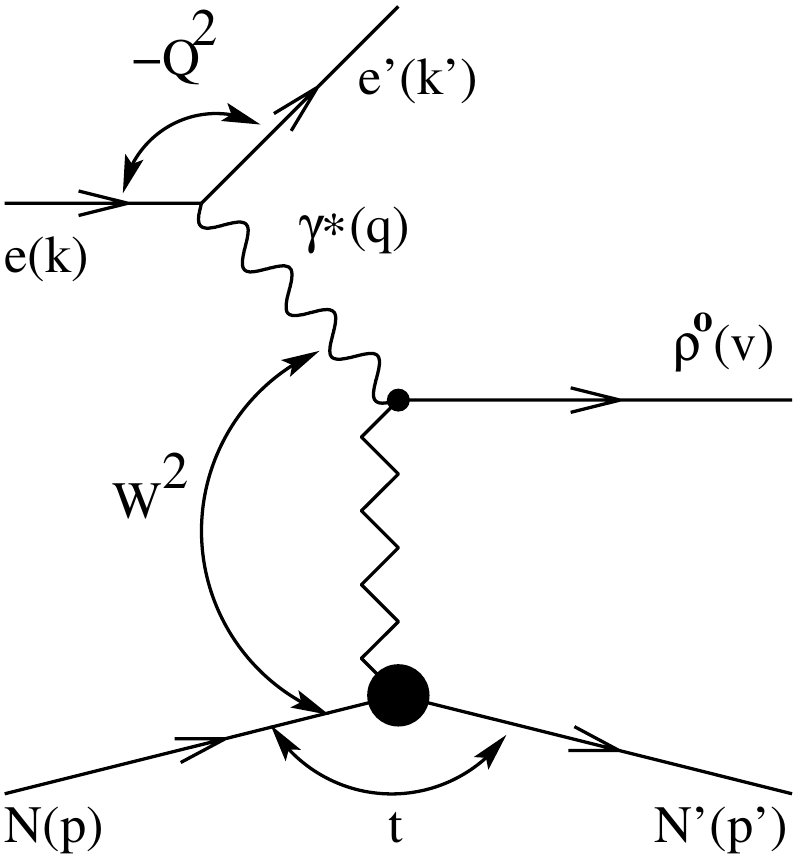}
\includegraphics[width=0.45\textwidth]{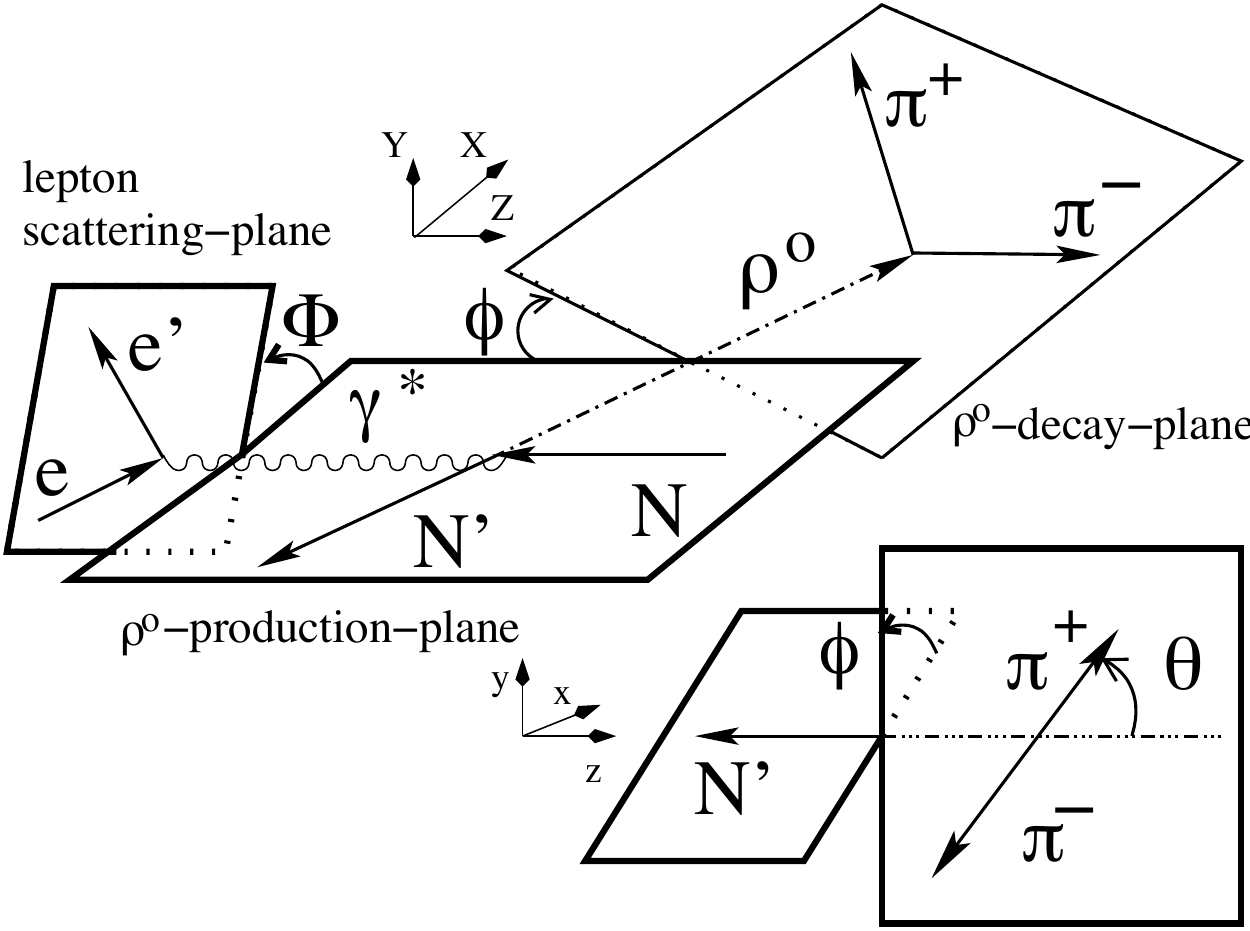}
\caption{Left: The electroproduction of a $\rho^0$ meson from a target proton. Right: The definition of the angles to which a fit is made to extract the helicity amplitude ratios.}
\label{p:angles}
\end{figure}
\end{center}

The helicity amplitude ratios can be extracted from the process $ep(d)\rightarrow ep(d)\rho^0\rightarrow ep(d)\pi^+\pi^-$ by fitting the decay pions from the $\rho^0$ meson with a 3D spatial distribution in the angles $\Phi$, which is the angle between the lepton scattering and $\rho^0$ production planes, $\phi$, which is the angle between the $\rho^0$ production and decay planes, and $\theta$, which is the angle in the $\rho^0$ rest frame between the 3-momentum of the decay $\pi^+$ and the direction opposite to that of the 3-momentum of the recoiling target. The process and the angular definitions are visualised in Fig.~\ref{p:angles}.

\section{The HERMES Experiment}
The H{\sc ermes} experiment~\cite{Ack98} has been covered in detail in the literature and no such detailed description will be repeated here. It is sufficient to note that the principle subdetectors in the system for the measurement under discussion were the tracking systems for the reconstruction of the scattered lepton and the gas threshold Cerenkov counter (until 1998) or the dual-radiator Ring-Imaging Cerenkov detector (after 1998). The Cerenkov detectors were used to identify the decay pions from the produced $\rho^0$ meson. The data were taken in 1996-2005 using unpolarised and polarised proton and deuteron targets, however the remaining average polarisation in the data set was negligible. Uncertainties arise in the measurements due to potential misidentification of the process because the recoiling proton is not detected and due to the H{\sc ermes} acceptance being $Q^2$-dependent. There is a third contribution to the systematic uncertainty due to the neglect of the amplitudes involving nucleon helicity flip and the sub-leading unnatural parity exchange amplitudes.

\section{Comparison with Previous Work}

The determination of the helicity amplitude ratios requires that those amplitudes relating to the flip of the nucleon helicity and sub-leading unnatural parity exchange are neglected. The resulting reduced parameter set allows the SDMEs to be calculated from the helicity amplitude ratios with, in some cases, better precision than the previous direct extraction. This is noticeably the case for the polarised SDMEs. The systematic uncertainty on each of the SDMEs includes a contribution due to the neglect of the amplitudes as described above. 

\section{Kinematic Dependences of Ratios}

\begin{center}
\begin{figure}
\includegraphics[width=0.45\textwidth]{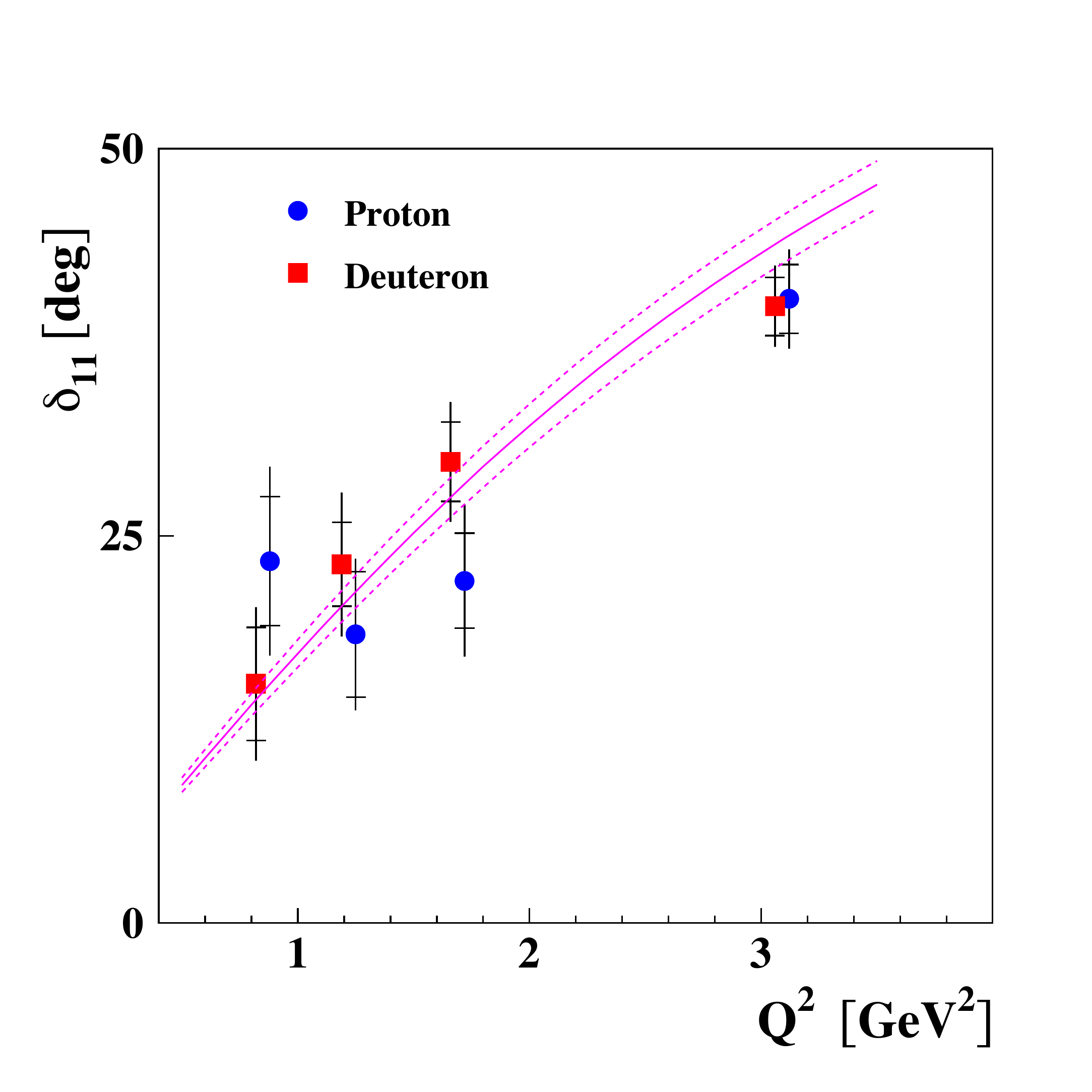}
\includegraphics[width=0.45\textwidth]{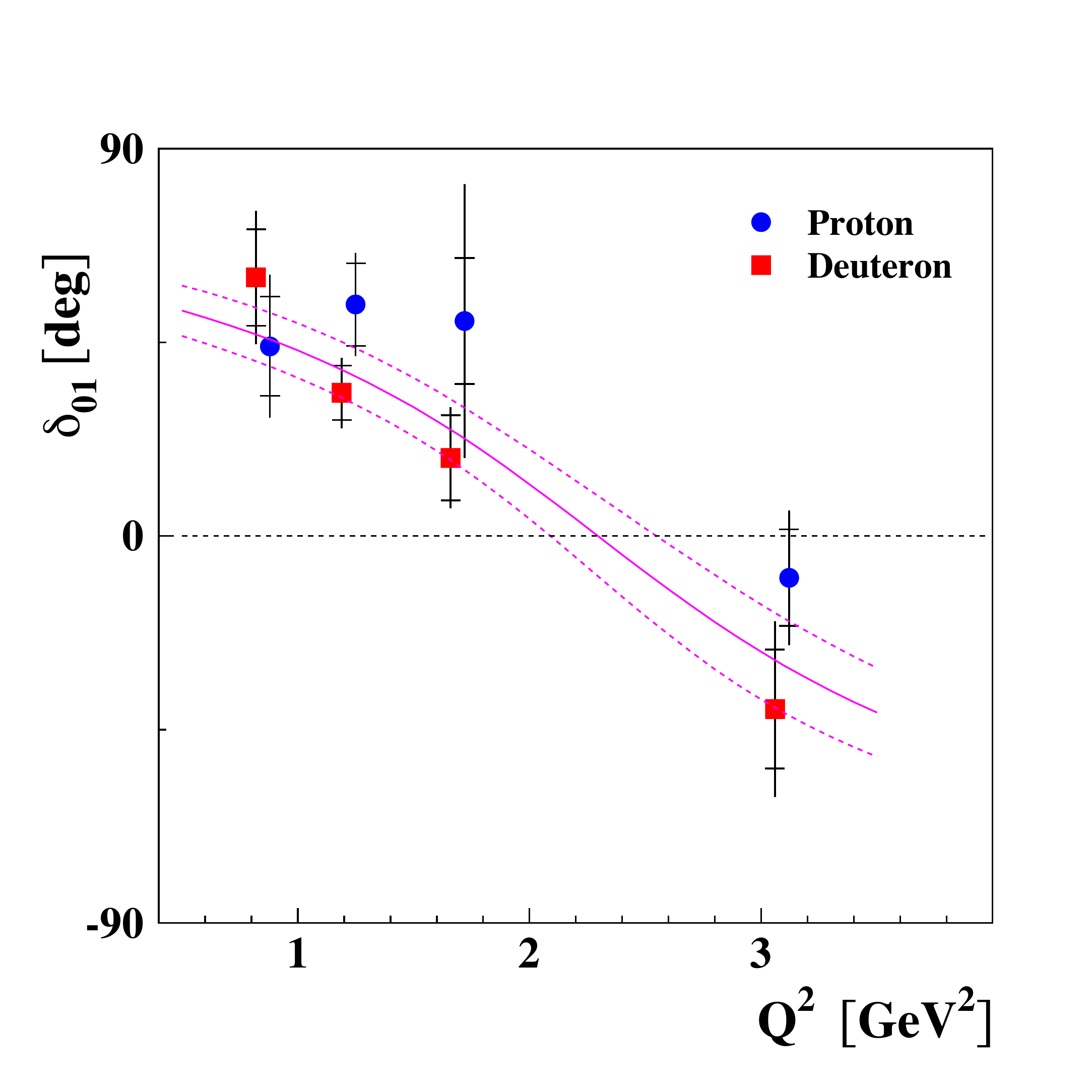}
\caption{The phase difference between the real and imaginary parts of the natural parity exchange amplitude ratios $T_{11}/T_{00}$ and $T_{01}/T_{00}$. The large phase differences that have a complicated dependance on $Q^2$ are unexpected. The displayed fits are from phenomenological fits that are not motivated by theory. See Ref.~\cite{paper} for more details.}
\label{p:phase}
\end{figure}
\end{center}

The kinematic dependences of the ratios can be calculated using assumptions from pQCD~\cite{ivanov,kuraev}. The ratio $T_{11}/T_{00}$ should follow a $1/Q$ dependence because $T_{11}$ is a twist-3 object and the expected behaviour can be seen in the real part of the ratio. However, the imaginary part of the ratio deviates from this behaviour and seems to increase linearly with $Q$. This leads the phase difference of the ratio to be large at larger $Q$ (see Fig.~\ref{p:phase}), which is unexpected and contradicts GPD models~\cite{gol1,gol2,gol3}. Unexpected kinematic behaviour is also displayed by the amplitude ratio $T_{01}/T_{00}$ which displays neither the expected behaviour in the real, nor in the imaginary part. Again, this leads to a phase difference that displays a complicated $Q^2$ dependence and contradicts simple pQCD expectations.

\section{Unnatural Parity Exchange}
 
\begin{center}
\begin{figure}
\includegraphics[width=0.45\textwidth]{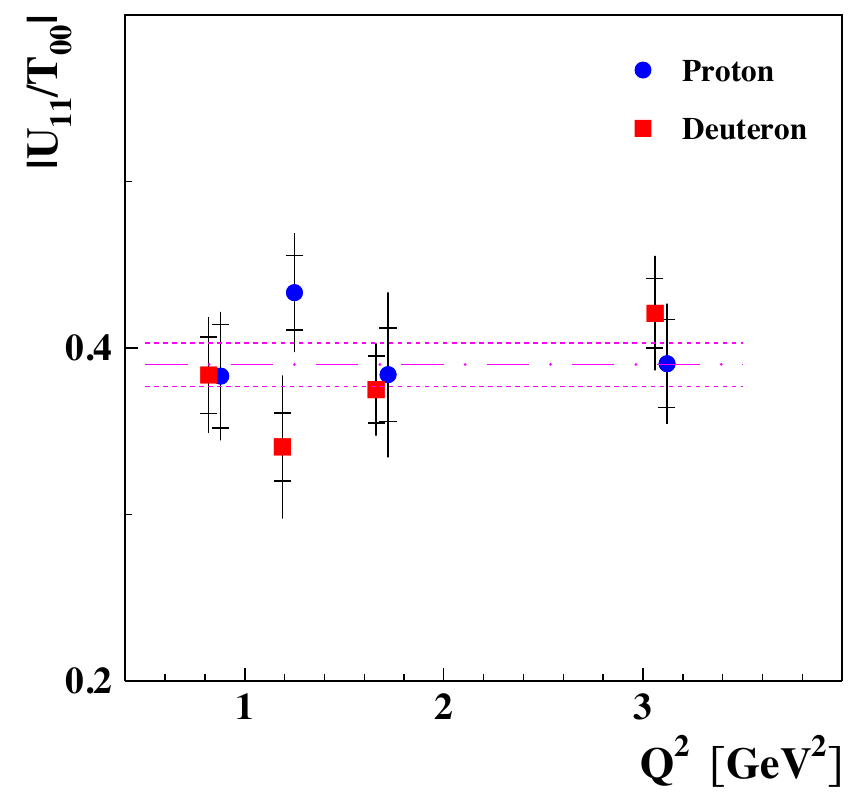}
\includegraphics[width=0.45\textwidth]{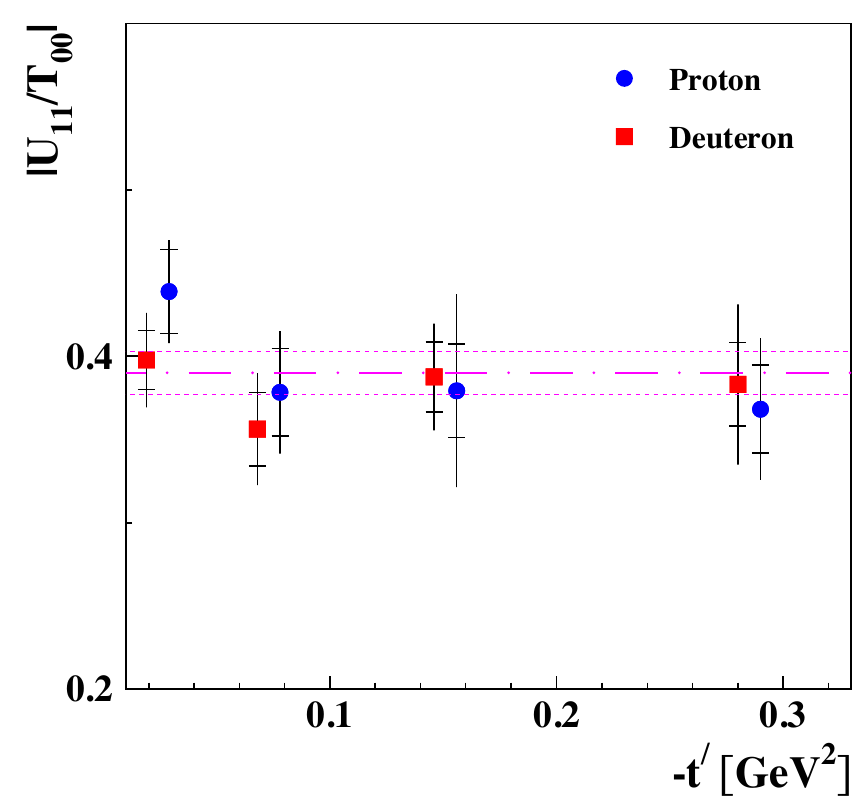}
\caption{The unnatural parity exchange signal $U_{11}/T_{00}$ projected in $Q$ (left) and $t'$ (right). The large signal is seemingly constant in both kinematic projections and its magnitude is greater than expected. The displayed fits are phenomenological fits to a constant.}
\label{p:upe}
\end{figure}
\end{center}

The $\rho^0$ meson is expected to display the largest contribution from unnatural parity exchange (UPE) processes of any meson at H{\sc ermes} kinematics as its quark content matches that from e.g. pion exchange. In the previous H{\sc ermes} publication on the SDMEs for $\rho^0$ electroproduction, the existence of UPE was confirmed to a level of 3~$\sigma$ of the uncertainty of the measurement. By combining the (consistent) data from proton and deuteron targets and extracting the UPE helicity amplitude ratio $U_{11}/T_{00}$ in a single kinematic bin, this existence is now confirmed to a level of 20~$\sigma$  of the experimental uncertainty. Furthermore, it is clear from the data that the UPE contribution has a value of $U_{11}/T_{00}$ of about 0.4 in any of four kinematic bins projected in $Q^2$ or $t'$ as shown in Fig.~\ref{p:upe}.

\section{Summary}

The helicity amplitude ratios relating to the process $ep\rightarrow ep\rho^0$ are extracted from a fixed target experiment for the first time. Comparisons with previously extracted SDMEs from the same data set shows that the amplitude method of analysis can recreate the SDMEs and often with better precision than achieved by direct extraction. The kinematic dependences shown by the amplitude ratios are often in contradiction to those expected from pQCD and contradict assumptions often inherent in GPD models. The signal relating to unnatural parity exchange is very strong and confirmed to 20 standard deviations. 

\bibliographystyle{aipproc}

\end{document}